\documentclass[pre,nofootinbib,twocolumn,floatfix,showpacs,sort&compress]{revtex4}
\usepackage{graphicx}
\usepackage{amsmath}
\usepackage{amssymb}

\ifnum\lefthyphenmin<2\lefthyphenmin=2\fi
\ifnum\righthyphenmin<2\righthyphenmin=2\fi

\begin{document}

\title{Denaturation of Circular DNA: Supercoils and Overtwist}

\author{Amir Bar$^{1,2}$, Alkan Kabak\c{c}{\i}o\u{g}lu$^{3}$, David Mukamel$^{1}$\emph{}\\
\emph{$^{1}$Department of Physics of Complex Systems and }\\
\emph{$^{2}$ Department of Computer Science and Applied Mathematics,
The Weizmann Institute of Science, Rehovot 76100, Israel}\\
\emph{$^{3}$Department of Physics, Ko\c c University, Sar\i yer 34450 \. Istanbul, Turkey}}
\begin{abstract}
The denaturation transition of circular DNA is studied within a
Poland-Scheraga type approach, generalized to account for the fact
that the total linking number (LK), which measures the
number of windings of one strand around the other, is conserved. In the model the LK
conservation is maintained by invoking both overtwisting and writhing (supercoiling) 
mechanisms. This generalizes previous studies which considered each mechanism
separately. The phase diagram of the model is analyzed as a function of the temperature
and the elastic constant $\kappa$ associated with the overtwisting energy for
any given loop entropy exponent, $c$. As is the case where the two mechanisms apply separately,
the model exhibits no denaturation transition for $c \le 2$. For $c>2$ and $\kappa=0$ we find that the
model exhibits a first order transition. The transition becomes
of higher order for any $\kappa>0$. We also calculate the contribution of
the two mechanisms separately in maintaining the conservation of the linking number
and find that it is weakly dependent on the loop exponent $c$.
\end{abstract}

\date{\today}

\pacs{87.15.Zg, 36.20.Ey}

\maketitle


\section{Introduction}

The thermal denaturation of DNA, whereby the two strands of the
molecule separate upon heating, has been thoroughly investigated both
experimentally and theoretically in the last half century. This
process is relevant for experiments such as polymerase chain
reaction (PCR) \cite{HM1994,HM1996}, and for biological processes such as those taking place within a thermophilic bacteria \cite{DSD1992,HS2004}. The
fraction of bound base pairs {\it vs.} temperature (the melting
curve) is measured by means of fluorescence and UV absorbtion methods.
A typical melting curve of chains of the order of thousands of base pairs is composed of a
sequence of discrete steps, interpreted as indicating a series of sharp, first-order
phase transitions corresponding to the local melting of regions with
successively increasing GC content.

A prototypical theoretical model for studying this phase transition is
the Poland-Scheraga (PS) model~\cite{PS1966}.
In this model, and for the case of a homopolymer DNA, the
molecule is represented by an alternating sequence of rigid bound
segments and flexible denatured loops. Their contribution to the
partition function are energetic and entropic, respectively. The
entropy $S(l)$ of a loop of length $l$ is of the form \[
e^{S(l)}\equiv\Omega(l) = A\,\frac{s^{l}}{l^{c}}\] where $A$ and $s$
are constants and $c$ is the \emph{loop exponent }
depending only on dimensionality and constraints imposed on the DNA
chain such as excluded volume interactions.
In the framework of the PS
model, the nature of the transition is set by the value of $c$: for
$c\le1$ no transition takes place and melting is just a gradual process in which the fraction of
bound base pairs is nonzero at all temperatures;
for $1<c\le2$ the transition is of second order;
for $c>2$ it is of first order, i.e., the melting curve is discontinuous at the melting temperature $T_c$.
It was shown relatively recently~\cite{KMP2000} that the excluded volume
corrections in three dimensions yield $c\approx2.12$. Therefore, the
PS model predicts a first-order melting transition.

The DNA is a double helix and in order to open a denatured loop the
region in which it is embedded must be unwound. This has no
consequence for a linear DNA chain in thermal equilibrium, where the
ends of the molecule are free to rotate. On the other hand, in
circular DNA (such as plasmids) and in DNA with rotationally fixed 
ends the total \emph{linking number} (LK), which measures the number of
windings of one strand about the other, is conserved. In such cases
the unwinding of one region must be compensated by the over-winding of
another region.

Two mechanisms have been suggested to absorb the extra linking number in the 
overwound regions: (a) increasing of twist
("Tw"), or \emph{overtwisting}, in which the change in the average stacking
angle accounts for the extra windings \cite{RB2002, GOY2004}, and (b)
increasing of writhe ("Wr"), or \emph{supercoiling}, where the backbone assumes 
a nonplanar shape that accommodates a nonzero LK \cite{YI2002,VAAP2000,KOM09,KOM2010,BKM2011}.
The C\u alug\u areanu-White-Fuller theorem implies that,
during the melting process one has $LK=Tw+Wr$ ~\cite{GC1959,GC1961,JHW1969,FBF1971}
It has 
been shown that the two mechanisms have similar effects on the melting
behavior: For $c\le2$, the melting process becomes a smooth crossover
with no phase transition. For $c>2$, there is a phase transition of
high order, where the singular part of the free energy scales with the reduced temperature $t\equiv(T-T_c)$ as $F_{sing}\sim
|t|^{(c-1)/(c-2)}$. Thus 
the order of the transition diverges as $c$ approaches $2$ from above,
and it becomes second order for $c\ge 3$.  Unlike the full
denaturation of DNA with free ends, the high-temperature phase here is
composed of a critical fluid of microscopic loops coexisting with a
single macroscopic loop \cite{BKM2011}.

In this paper, we investigate
the general scenario where both mechanisms act simultaneously. We find
that the nature of the transition is the same as that found in either
of the two mechanisms separately.
This observation is expected, although it is not
guaranteed by the fact that the two limits (supercoiling or
overtwisting alone) yield similar phase transition scenarios.
To probe the interplay between supercoils and overtwist we also calculate
the linking number absorbed by overtwist at the critical point. 

The paper is organized as follows: In section 2 the model is defined
and analyzed with a formalism somewhat different from earlier 
accounts.  In section 3 the results are presented, first for
simplified cases and then for the full model. We then conclude in
section 4 with a brief discussion of our results.

\section{The model}

\begin{figure}
\includegraphics[scale=0.35]{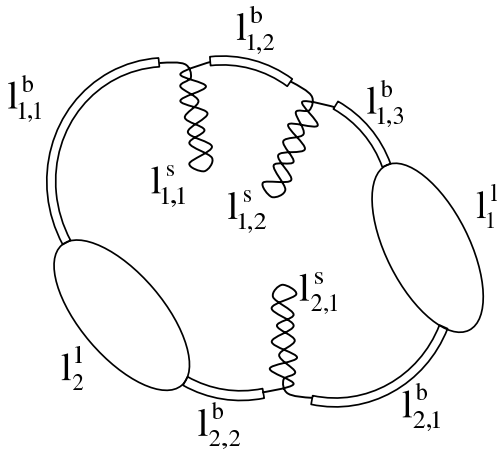}

\caption{\label{fig:model} A typical configuration of the model}

\end{figure}

In order to incorporate supercoils and overtwist,
we extend the PS model and assume that each configuration is composed
of an alternating sequence of bound segments, loops, and supercoils, the latter
being double stranded chains which carry writhe. 
It is assumed that supercoils form only within bound segments.
A typical configuration is sketched in Fig.\ref{fig:model}.
The contribution of the three types of segments to the free energy can be computed 
using the following rules:
\begin{itemize}
\item A bound segment of length $l$ contributes to the internal energy
  $E_{b}<0$ per unit length, and none to the entropy (due to the large persistence length of the 
  double stranded DNA). Hence, the associated Boltzmann weight is $e^{l\beta
    E_{b}}\equiv\omega^{l}$ with $\beta=1/k_{B}T$.
\item A supercoil of length $l$ contributes to the internal energy
  $E_{s}<0$ per unit length with $E_{s}>E_{b}$, yielding a Boltzmann weight $e^{l\beta
    E_{s}}\equiv\nu^{l}$. It is assumed that like the bound segments, these segments carry no internal entropy
    although they do contribute to the overall entropy through
    their positional degree of freedom.
\item A loop of size $l$ has an entropic contribution given by
  the Boltzmann weight $\Omega(l)=A\frac{s^{l}}{l^{c}}$. Here $s$ is a geometry dependent constant
  and $A$ is a constant,
  usually termed the {\it cooperativity parameter}, reflecting both the
  normalization of the entropic contribution and the enthalpic cost of
  initiating a new loop.
\end{itemize}
In addition, we assume an overtwisting elastic energy cost with an elastic constant $\kappa$.

The excess linking number residing on the double-stranded DNA segments
(bound segments and supercoils) is calculated as follows: A unit
length of a loop region increases LK by $1$, while a supercoil segment of the same length
decrease LK by $1$. The $1:1$ ratio is assumed for the sake of
simplicity, while considerations of universality suggest that the results
should remain qualitatively unaltered under different
choices. Denoting by $L_b, L_s$ and $L_l$ the total length of the
bound segments, supercoils and loops, respectively, the
excess LK in the double stranded regions is simply $L_{l}-L_{s}$. This is
compensated by an increase in the average stacking angle per unit length by
$\Delta\theta$ in the bound segments and supercoils combined, hence
$\Delta\theta=\frac{L_{l}-L_{s}}{L_{b}+L_{s}}$ \cite{RB2002}. Then,
the elastic energy cost due to overtwisting is
$\kappa\left(L_{b}+L_{s}\right)\left(\Delta\theta\right)^{2}=\kappa\frac{\left(L_{l}-L_{s}\right)^{2}}{L_{b}+L_{s}}$,
yielding the Hamiltonian\[
H=L_{b}\epsilon_{b}+L_{s}\epsilon_{s}+\kappa\frac{\left(L_{l}-L_{s}\right)^{2}}{L_{b}+L_{s}}\]
Thus, for example, the Boltzmann weight of the configuration depicted
in Fig.\ref{fig:model} is given by
\begin{eqnarray*}
&& \omega^{l^b_{1,1}}\nu^{l^s_{1,1}}\omega^{l^b_{1,2}}\nu^{l^s_{1,2}}\omega^{l^b_{1,3}}\Omega\left(l^l_1\right)\omega^{l^b_{2,1}}\nu^{l^s_{2,1}}\omega^{l^b_{2,2}}\Omega\left(l^l_2\right) \\
&& \qquad \times \; e^{-\beta\kappa\frac{\left(l^l_1+l^l_2-l^s_{1,1}-l^s_{1,2}-l^s_{2,1}\right)^2}{L-l^l_1-l^l_2}}
\end{eqnarray*}
It is worth noting that some of the previously studied PS-type models can
be formulated as special cases of the present model: $L_{s}=0$ corresponds to the case
with overtwisting only \cite{RB2002}; $L_{s}=L_{l}$ corresponds to the
case with supercoils only \cite{KOM09,BKM2011} and $\kappa=0$ corresponds
to a DNA with supercoils and no LK constraint \cite{KOM09}.

The canonical partition function can now be written as
\begin{equation}
Z(L)=\sum_{L_{b}+L_{l}+L_{s}=L}Z_{\kappa=0}(L_{b},L_{l},L_{s})e^{-\beta\kappa\frac{\left(L_{l}-L_{s}\right)^{2}}{L_{b}+L_{s}}},\label{eq:canonical_Z}\end{equation}
where $Z_{\kappa=0}(L_{b},L_{l},L_{s})$ is the partition sum with
given $L_{b,l,s}$ and $\kappa=0$ (which is an ensemble more restricted than
even the microcanonical ensemble, as there may be different
$L_{b,l,s}$ triplets that have the same energy).  The correspondence
with other models mentioned above can be obtained from
Eq.(\ref{eq:canonical_Z}) by taking the appropriate limits: The PS
model is recovered when $\kappa=0$ and $\nu=0$ (or $E_{s}=\infty$ so
that $L_{s}=0$); taking $\nu=0$, $\kappa>0$ yields the partition sum for
a model with overtwisting only \cite{RB2002}; finally, substituting
$\kappa=\infty$, $\nu>0$ recovers the case with supercoils only
\cite{KOM09}.


\subsection{Free energy}

We begin by calculating $Z_{\kappa=0}(L_{b},L_{l},L_{s})$. This is conveniently done
by first evaluating the grand canonical partition sum by means of a
$z$-transform of $Z_{\kappa=0}$. The canonical partition sum
(expressed in terms of $L_{b},L_{l}$ and $L_{s}$, or in terms of the
fractions $m_{i} \equiv L_{i}/L$, $i=b,s,l$ ) is then calculated using the inverse
transform. Introducing three fugacities, $z_{b},z_{l}$, and
$z_{s}$, corresponding to the three length constraints, the resulting
grand canonical partition function $Q_{\kappa=0}(z_{b},z_{s},z_{l})$
can be expressed in a closed form as
\begin{eqnarray} Q_{\kappa=0}(z_{b},z_{s},z_{l}) & = &
\sum_{L_{b,s,l}}Z_{\kappa=0}\left(L_{b},L_{s},L_{l}\right)z_{b}^{L_{b}}z_{s}^{L_{s}}z_{l}^{L_{l}}\nonumber
\\ & = &
1+\tilde{V}\left(z_{b},z_{s}\right)U\left(z_{l}\right)+\nonumber \\ &
&
+\tilde{V}\left(z_{b},z_{s}\right)U\left(z_{l}\right)\tilde{V}\left(z_{b},z_{s}\right)U\left(z_{l}\right)
\nonumber \\ && + \cdots \nonumber \\ &
= &
\frac{1}{1-\tilde{V}\left(z_{b},z_{s}\right)U\left(z_{l}\right)},\label{eq:Q_exp1}\end{eqnarray}
with
\begin{eqnarray} U\left(z_{l}\right) & = &
A\sum_{l=1}^{\infty}\frac{s^{l}z^{l}}{l^{c}}=A\Phi_{c}\left(sz_{l}\right), \\
\tilde{V}\left(z_{b},z_{s}\right)
& = &
V\left(z_{b}\right)+V\left(z_{b}\right)W\left(z_{s}\right)V\left(z_{b}\right)+... \nonumber \\ &
= &
\frac{V\left(z_{b}\right)}{1-W\left(z_{s}\right)V\left(z_{b}\right)}, \\
V\left(z_{b}\right)
& = & \sum_{l=1}^{\infty}\omega^{l}z_{b}^{l}=\frac{\omega
  z_{b}}{1-\omega z_{b}},\\
  W\left(z_{s}\right) & = &
\sum_{l=1}^{\infty}\nu^{l}z_{s}^{l}=\frac{\nu z_{s}}{1-\nu
  z_{s}}.\label{eq:W_def}
\end{eqnarray}
The functions $U\left(z_{l}\right)$ and $\tilde{V}\left(z_{b},z_{s}\right)$ are the
grand canonical sums of single-stranded (loops) and double-stranded
(bound and supercoiled) segments, respectively. Similarly,
$V\left(z_{b}\right)$ and $W\left(z_{s}\right)$ denote the grand sums
for bound and supercoiled segments, separately.  Here
$\Phi_{c}\left(q\right)$ is the polylogarithm function of order $c$,
which is analytic everywhere except for a branch-cut for $q\ge1$.  The
behavior of this function at $q=1$ depends on $c$: If $c\le1$, $\Phi_c(q)$ diverges as
$q\rightarrow1^{-}$. If $c>1$, $\Phi_{c}(q\rightarrow1^-)=\zeta_{c}$
where $\zeta_{c}$ is the Riemann zeta function \cite{AS1964}. The
behavior of $\Phi_{c}(q)$ near $q=1$ determines the nature of the
phase transition investigated here, as will be shown below.
In deriving Eq.(\ref{eq:Q_exp1}) we take $Z_{\kappa=0}(0,0,0)=1$ and
assume that the chain contains at least one loop and one bounded segment.
This assumption simplifies the numerator of the resulting expression in (\ref{eq:Q_exp1})
and it has no effect on the resulting thermodynamic properties of the model.

The canonical partition function is found by inverting the z-transform using a Cauchy integral:
\begin{equation}
Z_{\kappa=0}(L_{b},L_{l},L_{s})=\left(\frac{1}{2\pi
  i}\right)^{3}\oint\frac{Q_{\kappa=0}(z_{b},z_{s},z_{l})}{z_{b}^{L_{b}+1}z_{s}^{L_{s}+1}z_{l}^{L_{l}+1}}dz_{b}dz_{s}dz_{l}\ .\label{eq:Z_cauchy}\end{equation}
All integration contours encircle the origin and contain no other
singularities. Using Eqs.(\ref{eq:Q_exp1}-\ref{eq:W_def}) we find \[
Q_{\kappa=0}(z_{b},z_{s},z_{l})=\left[\frac{1}{\omega
    z_{b}}-\frac{1}{1-\nu
    z_{s}}-A\Phi_{c}\left(sz_{l}\right)\right]^{-1}.\] 
 $Q_{\kappa=0}$ has a simple pole in $z_b$ set by
\begin{equation}
  \frac{1}{\omega z_{b}}-\frac{1}{1-\nu
    z_{s}}-A\Phi_{c}\left(sz_{l}\right)=0, \label{eq:zb_eq}
\end{equation}
yielding
\begin{equation}
      z_{b}^{*} = \frac{1}{\omega}\left[\frac{1}{1-\nu
      z_{s}}+A\Phi_{c}\left(sz_{l}\right)\right]^{-1}.\label{eq:zb_sol}
\end{equation}
Note that Eq.(\ref{eq:zb_eq}) is equivalent to Eq.(2) in \cite{BKM2011}. The $z_b$
contour in Eq.(\ref{eq:Z_cauchy}) can be deformed to encircle the pole given by
Eq.(\ref{eq:zb_sol}), yielding
\begin{equation}
  Z_{\kappa=0}\left(m_{b},m_{l}\right)=\left(\frac{1}{2\pi
    i}\right)^{2}\oint
  e^{-L\tilde{F}_{\kappa=0}\left(z_{s},z_{l},m_{b},m_{l}\right)}dz_{l}dz_{s}\label{eq:Z_fe}\end{equation}
with \begin{eqnarray}
  \tilde{F}_{\kappa=0}\left(z_{s},z_{l},m_{b},m_{s}\right) & = &
  m_{b}\log\left[z_{b}^{*}\left(z_{s},z_{l}\right)\right]+m_{s}\log\left(z_{s}\right)\nonumber
  \\ & &
  +\left(1-m_{b}-m_{s}\right)\log\left(z_{l}\right) \label{eq:F_k=00003D0_def}\end{eqnarray}
up to logarithmic corrections in $L$. Here we used the fact that
$m_{b}+m_{s}+m_{l}=1$. In the thermodynamic limit the integral in
Eq.(\ref{eq:Z_fe}) can be evaluated by considering the saddle point of
$\tilde{F}_{\kappa=0}$ with respect to $z_{l}$ and $z_{s}$,
\begin{eqnarray} 0 &
  = & \frac{\partial\tilde{F}_{\kappa=0}}{\partial z_{s}}=-\frac{m_{b}\nu/(1-\nu
    z_{s})}{1+(1-\nu
    z_{s})A\Phi_{c}(sz_{l})}+\frac{m_{s}}{z_{s}}\ ,\label{eq:dFdzs}\\ 0
  & = & \frac{\partial\tilde{F}_{\kappa=0}}{\partial z_{l}}=-\frac{m_{b}(1-\nu
    z_{s})A\Phi_{c-1}(sz_{l})}{z_{l}\big[1+(1-\nu
      z_{s})A\Phi_{c}(sz_{l})\big]}\nonumber \\ &&\ \ \ \ \ \ \ \ \ \ \ \ \ \ +
  \frac{1-m_{b}-m_{s}}{z_{l}}\label{eq:dFdzl}\end{eqnarray} where we
used the identity
$\frac{d}{dq}\Phi_{c}\left(q\right)=\frac{1}{q}\Phi_{c-1}\left(q\right)$.
After some algebra Eq.(\ref{eq:dFdzs}) yields
\begin{equation}
  z_{s}^{*}=\frac{1}{\nu}\left[1+\frac{1+x-\sqrt{\left(1+x\right)^{2}+4x\eta}}{2\eta}\right],\label{eq:zs_sol}
\end{equation}
where we define
\begin{equation}
x\equiv \frac{m_{b}}{m_{s}}\quad,\quad \eta\equiv
A\Phi_{c}(sz_{l}). \label{eq:xandeta}
\end{equation}
It can be seen (as $x,\eta>0$) that
$z_{s}^{*}<\nu^{-1}$, therefore the $z_{s}$ integration contour can be 
deformed to pass through this saddle point without encircling the
singularity of $\tilde{F}$ at $z_{s}=\nu^{-1}$. Eq.(\ref{eq:dFdzl})
yields
\begin{equation}
  1+\frac{A\Phi_{c-1}(sz_{l})}{\frac{1}{1-\nu
      z_{s}^{*}}+A\Phi_{c}(sz_{l})}=\frac{1}{m_{b}}-\frac{1}{x}.\label{eq:zl_eq}
\end{equation}
The LHS of Eq.(\ref{eq:zl_eq}) is monotonically increasing with $z_{l}$
(see Appendix A). For $c\le2$ this equation has a solution,
$z_{l}^{*}$, for any value of $m_{b}$ and $x$ due to the fact that $\Phi_{c-1}(sz_{l})$ diverges at $sz_l=1$. However, for $c>2$, 
the LHS reaches a finite value for $z_{l}=s^{-1}$
(the branch point of $\Phi_c(sz_l)$). Therefore, for a given value of $x$,
there is no saddle point for values of $m_{b}$ below a critical
threshold $m_{b}^{(c)}$ given by
\begin{equation}
m_{b}^{(c)}(x)=\left[1+\frac{1}{x}+\frac{A\zeta_{c-1}}{\frac{1}{1-\nu
      z_{s}^{*}}+A\zeta_{c}}\right]^{-1}.\label{eq:mb_c_vs_x}
\end{equation}
Here $z_{s}^{*}$ is obtained using Eq.(\ref{eq:zs_sol}) with $\eta$ replaced by $A\zeta_{c}$.
For values $m_{b}<m_{b}^{(c)}(x)$ the $z_{l}$
integral is equal to the value of the integrand at
the branch point $z_{l}=s^{-1}$, as in the canonical treatment of the
PS model \cite{Wie1983}. The integration procedure involves more 
than simply evaluating the integrand at the
singularity closest to the origin. Details are given in Appendix B.

To calculate the canonical partition function $Z(L)$ given in Eq.(\ref{eq:canonical_Z}) the overtwist term should be added to the free energy, yielding
\begin{eqnarray}
\tilde{F}\left(z_{s},z_{l},m_{b},m_{s}\right) &
  = & m_{b}\log\left[z_{b}^{*}\left(z_{s},z_{l}\right)\right]\nonumber
  \\ & + & 
  m_{s}\log\left(z_{s}\right)+\left(1-m_{b}-m_{s}\right)\log\left(z_{l}\right)\nonumber
  \\ & + & 
  \beta\kappa\frac{\left(1-m_{b}-2m_{s}\right)^{2}}{m_{b}+m_{s}}\ .\label{eq:F_def}
\end{eqnarray}
This full free energy needs to be minimized with respect to all of its arguments. The minimization with respect to $z_s$ and $z_l$ is the same as for $\tilde{F}_{\kappa=0}$ and the results are given by Eq.(\ref{eq:zs_sol},\ref{eq:zl_eq}). The minimization with respect to $m_{b}$ and $m_{s}$ is discussed in the next section.


In summary, the fugacities $z_{s},z_{l}$ in the thermodynamic limit 
(denoted by $z_{s}^{*}$ and $z_{l}^{*}$) as functions of the bound and supercoiled segment fractions, $m_{b}$ and $m_{s}$, are given by
Eqs.(\ref{eq:zs_sol},\ref{eq:zl_eq}). Hence we can express the Landau
free energy, Eq.(\ref{eq:F_def}), as a function of the densities $m_{b}$ and $m_{s}$ only. In what follows it will be occasionally more convenient to express the dependence on $m_s$ through the fraction
$x=m_{b}/m_{s}$ as defined in Eq.(\ref{eq:xandeta}).
Then the Landau free energy can be written as
\begin{eqnarray}
F\left(m_{b},m_{s}\right) & = &
F_{\kappa=0}\left(m_{b},x\right)+\beta\kappa\frac{\left(1-m_{b}-2m_{s}\right)^{2}}{m_{b}+m_{s}}, \nonumber \\
&& \label{eq:redF_def}\\ F_{\kappa=0}\left(m_{b},x\right)
& = &
\tilde{F}_{\kappa=0}\left(m_{b},m_{s},z_{l}\left(m_{b},x\right),z_{s}\left(m_{b},x\right)\right)\nonumber
\\ & = & m_{b}\log\left[z_b(z_s,z_l)\right] + \frac{m_{b}}{x}\log\left(z_{s}\left(x,z_{l}\right)\right) \nonumber
\\ & + & (1-m_{b}\frac{x+1}{x})\log\left(z_{l}\right), \label{eq:redF_k=00003D0}
\end{eqnarray}
where $z_b$ is given by Eq.(\ref{eq:zb_sol}), $z_s$ is given by Eq.(\ref{eq:zs_sol}) and $z_{l}=z_{l}\left(m_{b},x\right)$ is given by Eq.(\ref{eq:zl_eq}).
Note that unlike other Landau free energies which are analytic in the order parameter, here $F_{\kappa=0}(m_b,x)$ is a non-analytic function of $m_b$ and $x$ along the line defined by $m_{b}^{(c)}(x)$. We shall denote this line of singularities by $\Gamma$. In the $m_b-m_s$ plane the expression for $\Gamma$ is
\begin{equation}
\Gamma: \left(m_b,m_s\right)=\left(m_{b}^{(c)}(x) , \frac{m_{b}^{(c)}(x)}{x}\right)\ \ ; \ \ x\in (0,\infty). \label{eq:Gamma} 
\end{equation}

For points to the left of $\Gamma$ (i.e. those for which $m_{b}<m_{b}^{(c)}(x)$ so that
$z_{l}=s^{-1}$) and for a given $x$ the free energy is linear in $m_b$,
while for points above $\Gamma$ it has a more complicated form, hence
$F_{\kappa=0}(m_b,m_s)$ is singular along $\Gamma$. Below we will explore this non-analyticity in more detail and show that it is closely related to the non-analytic behavior of the free energy as a function of temperature at the transition point.

\section{Results}
\label{sec:results}
After introducing the Landau free energy $F(m_{b},m_{s})$ and arguing
that it is singular on a line $(\Gamma)$ in the
$\left(m_{b},m_{s}\right)$ plane, we move on to study the nature
of the phase transition for different values of $\kappa$. Three cases
are of interest:
\begin{itemize}
\item $\kappa=0$: Here overtwisting has no cost and the chain is
  equivalent to a linear chain with supercoils freely spread within,
  with no linking number constraint. We will see that in this case the
  transition (which exists only for $c>2$) is first order as in the
  standard PS model.
\item $\kappa=\infty$: Here overtwisting is forbidden. This is the case
  with supercoils only which was analyzed in Ref.\cite{BKM2011} and
  found to exhibit a continuous transition of order $\left\lceil
  \frac{c-1}{c-2}\right\rceil $ with a singularity in the free energy which scales as $\sim t^{(c-1)/(c-2)}$.
  Here we will outline the derivation
  of this result within the current approach.
\item $0<\kappa<\infty$: In this case supercoiling and overtwisting coexist,
  yielding a different free energy minimum. Yet, it is shown that the
  nature of the transition remains the same as for $\kappa=\infty$.
\end{itemize}
To quantify the interplay between supercoiling and overtwisting, we
calculate the fraction of the linking number accommodated by overtwist at
the transition point
\begin{equation}
r(\kappa)\equiv m_{lc} - m_{sc} = 1-m_{bc}-2m_{sc}.\label{eq:r_def}
\end{equation}
Clearly, $r(\kappa=\infty)=0$ as no overtwist is allowed in this
limit. Below we derive an explicit formula for $r(\kappa=0)$ and
calculate  $r(0<\kappa\le\infty)$ numerically.

Throughout the paper the parameters used in the figures are
$E_{b}=-3$, $E_{s}=-2$, $s=5$, $A=0.1$.

\subsection{$\kappa=0$ }

The densities $m_{s}$ and $m_{b}$ are found by minimizing
$F\left(m_{b},x\right)\equiv F_{\kappa=0}\left(m_{b},x\right)$. 
This is equivalent to minimizing
$\tilde{F}\left(z_{s},z_{l},m_{b},m_{s}\right)$ with respect to all of
its arguments, as $F$ is obtained from $\tilde{F}$ by minimizing it with
respect to the fugacities $z_{s}$ and $z_{l}$. Using
Eq.(\ref{eq:redF_k=00003D0}) and minimizing $F\left(m_{b},m_{s}\right)$
yields
\begin{equation}
  z_{s}\left(m_{b},x\right)=z_{l}\left(m_{b},x\right)=z_{b}\left(m_{b},x\right),\label{eq:case_k=00003D0}
\end{equation}
i.e., the system is described by a single fugacity. This is expected,
since for $\kappa=0$ the grand canonical partition function of the full model could have been derived with a single fugacity corresponding to the single constraint $L_{b}+L_{s}+L_{l}=L$.

Substituting Eq.(\ref{eq:case_k=00003D0}) in
Eq.(\ref{eq:redF_k=00003D0}) one finds
\[
F\left(T\right)=\log\left[z_{l}\left(m_{b}(T),x(T)\right)\right],
\]
where $m_b$, $m_s$ and $z_l$ can be calculated by solving (\ref{eq:zb_sol},\ref{eq:zs_sol},\ref{eq:zl_eq}) using Eq.(\ref{eq:case_k=00003D0}). Hence the non-analytic behavior of $z_{l}$ results in a singularity in
$F$. As mentioned above, for $c\le2$, $z_{l}\left(m_{b},m_{s}\right)$
is an analytic function, therefore there is no phase transition in the
system. For $c>2$, $z_{l}$ increases monotonically from zero with
temperature as long as $T<T_{c}$, where using Eq.(\ref{eq:zb_sol}) the critical temperature 
$T_{c}$ is given by
\[
\frac{1}{\omega\left(T_{c}\right)}\left[\frac{1}{1-\nu\left(T_{c}\right)/s}+A\zeta_{c}\right]^{-1}=\frac{1}{s}.\]
For $T>T_{c}$, $z_{l}=s^{-1}$ is a constant and hence $F(m_b,x)$ is constant, independent of $T$. As $T\rightarrow T_c$ from below, the free energy approaches the transition point with a non-zero slope, i.e., the
transition is first order. This can be seen by differentiating Eq.(\ref{eq:zb_sol}) with respect to $T\sim \log(\omega)$ using relation (\ref{eq:case_k=00003D0}). 

\begin{figure}
\includegraphics[scale=0.5]{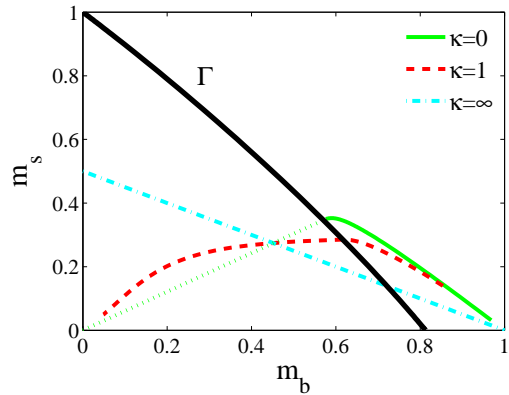}

\caption{\label{fig:mbms_plane} (Color online) The $(m_{b},m_{s})$ trajectories as a
  function of temperature of  for $\kappa=0$ (green
  full and dotted line), $\kappa=1$ (red dashed line) and
  $\kappa=\infty$ (cyan dash-dotted line). For all lines $c=2.5$. At $T=0$ the chain is
  totally bound, so $m_{b}=1$ and $m_{s}=0$ for all trajectories. As
  $T$ increases $m_{b}$ decreases, intersecting at $T=T_{c}$ the
  singular line $\Gamma$ (Eq.(\ref{eq:Gamma})), drawn
  above as a thick black line. On the $\kappa=0$ trajectory there
  is a coexistence between a bound phase and a denaturated phase at
  $T=T_{c}$, along the dotted green line, while for $T>T_{c}$ the system is effectively
  unbound with $m_{b}=m_{s}=0$.  For $\kappa=1$ (and any $\kappa>0$)
  the trajectory continues smoothly across the the singular line and
  reaches $m_{b}=m_{s}=0$ only at $T=\infty$. For $\kappa=\infty$ the
  trajectory is linear due to the simple relation
  $m_{s}=\frac{1}{2}(1-m_{b})$.
   }
\end{figure}

The equilibrium values of $m_b$ and $m_s$ as a function of $T$ define a trajectory in the $m_b-m_s$ plane, as depicted in Fig.\ref{fig:mbms_plane}. The starting point of this trajectory at $T=0$ is to the right of the singular line $\Gamma$ since $m_b>m_b^{(c)}(x)$. As $T$ increases, $m_b$ decreases. At $T=T_c$ where the trajectory intersects $\Gamma$, the singular line defined above, a phase transition takes place.

\begin{figure}
\includegraphics[scale=0.6]{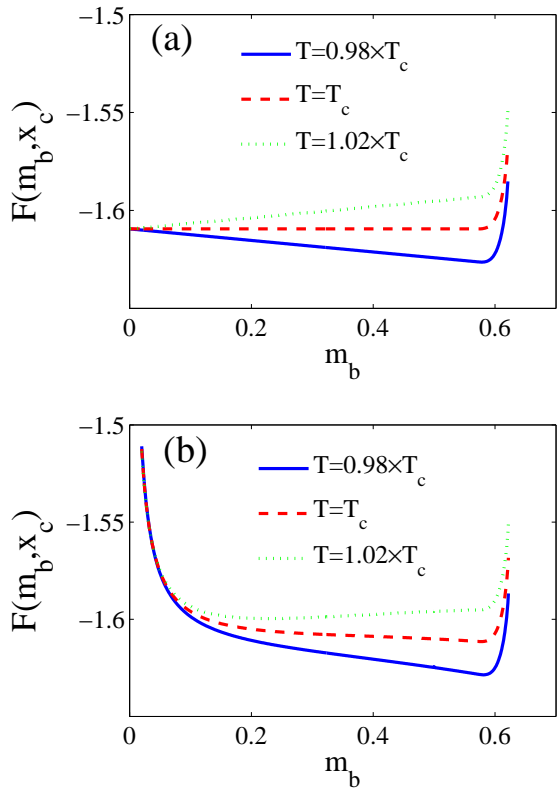}

\caption{\label{fig:F_vs_mb}The Landau free energy $F(m_{b},m_{s})$ as
  defined in Eq.(\ref{eq:redF_def}) along the line $m_{b}=x_c m_{s}$
  where $x_c=\frac{m_{bc}}{m_{sc}}$, for $c=2.5$ and (a) $\kappa=0$ ; (b)
  $\kappa=0.01$. While the free energy at the phase transition point for $\kappa=0$ has a continuum of minima in the interval $m_{b}<m_{b}^{(c)}(x)$, for $\kappa=0.01$
  (and for any $\kappa>0$) there is a unique minimum of the free
  energy at all temperatures.}

\end{figure}

The fact that the transition is first order can be verified as follows: the intersection of the trajectory with $\Gamma$ takes place at a certain $x_{c}=m_{bc}/m_{sc}$ where $m_{bc}$ and $m_{sc}$ are the critical fractions of the bound segments and the supercoils on $\Gamma$, respectively. Since the minimum of the Landau free energy at the critical temperature is obtained at $\left(m_{bc},m_{sc}\right)$, the slope of this free energy vanishes in all directions, i.e. $\partial F\left(m_{bc},x_c\right)=0$. As stated above and can been seen by inspecting Eq.(\ref{eq:redF_k=00003D0}), for points to the left of $\Gamma$, where $m_b<m_{bc}$ and $T>T_c$, the free energy $F\left(m_b,x\right)$ is linear in $m_b$ for fixed $x$. Hence the slope of $F\left(m_b,x_c\right)$ for $m_b<m_{bc}$ must be $0$. This implies, in particular, that $F\left(m_{bc},x_c\right) = F\left(0,x_c\right)$, namely a phase coexistence between bound and unbound phases. Note that there is no free energy barrier between the two phases.
As depicted in Fig.\ref{fig:F_vs_mb}(a) and can be verified by Eq.(\ref{eq:redF_k=00003D0}) above $T_c$ the slope of $F\left(m_b,x_c\right)$ for $m_b<m_{bc}$ is positive and hence $m_b=m_s=0$ is the minimal solution, so the system is in the unbound phase. As will be discussed below, setting
$\kappa>0$ eliminates the coexistence and yields a unique free energy
minimum at all temperatures (see Fig.\ref{fig:F_vs_mb}(b)).

Let us now consider the overtwist linking number $r(\kappa=0)$.
The value of
$x=m_{b}/m_{s}$ at criticality can be calculated 
using Eqs.(\ref{eq:dFdzs},\ref{eq:case_k=00003D0}):
\begin{eqnarray*}
x_{c}^{-1} & = & \frac{\nu s^{-1}/\left(1-\nu s^{-1}\right)}{1+(1-\nu
  s^{-1})A\zeta_{c}}.\end{eqnarray*} Solving Eq.(\ref{eq:mb_c_vs_x})
and $m_{sc}=m_{bc}x_{c}^{-1}$ for $m_{bc}$ and $m_{sc}$, we obtain the
overtwist linking number at $T_c$ as 
\[
r(\kappa=0)=\frac{A\zeta_{c-1}\left(1-\nu s^{-1}\right)^{2}-\nu s^{-1}}{1 + \left(1-\nu s^{-1}\right)^{2}\left(A\zeta_{c}+A\zeta_{c-1}\right)}\ .
\]
Depending on parameters in this expression, $r(\kappa=0)$ can be either positive or negative. Specifically, for the parameters used in Fig.\ref{fig:mbms_plane} the value is $r(\kappa)=-0.26343<0$, implying that, the length of the supercoiled regions at the phase transition point 
exceeds the length needed to compensate for the linking number released by the loops, resulting in undertwisted bound segments. 

\subsection{$\kappa=\infty$ }

\begin{figure}
\includegraphics[scale=0.5]{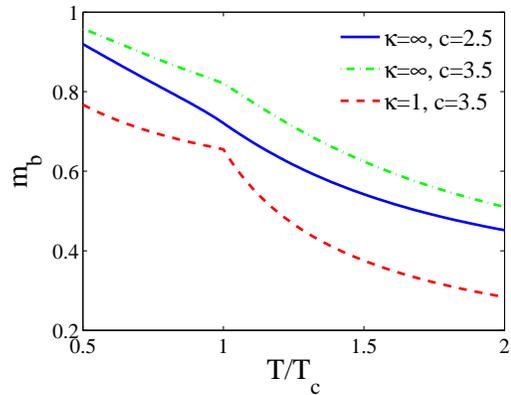}

\caption{\label{fig:mb_vs_T}$m_{b}$ vs. $T$ for various $c$ and
  $\kappa$.  The signature of the second order transition is the non
  differentiability of the curves at $T=T_{c}$ when $c>3$.  For $2<c<3$
  the melting curve is smooth with a higher order singularity at $T=T_c$.
  It can also be seen that the transition sharpens as $\kappa$ decreases.}

\end{figure}

When $\kappa=\infty$, overtwisting is forbidden. The conservation of
the linking number now implies $m_{s}=m_{l}$, therefore
$m_{s}=\frac{1-m_{b}}{2}$.  Minimizing Eq.(\ref{eq:F_def})
with respect to $m_{b}$ and using the linking number constraint yields
the relation\begin{equation}
z_{b}=\sqrt{z_{s}z_{l}},\label{eq:case_k=00003Dinf}\end{equation}
which implies that two fugacities are needed, accounting for the two
constraints on the linking number and the total chain length. Indeed, in
previous accounts of this model the derivation was conducted 
using two fugacities 
\cite{KOM09,BKM2011}.  Inserting Eq.(\ref{eq:case_k=00003Dinf}) into
Eq.(\ref{eq:F_k=00003D0_def}) yields \begin{equation}
  F=\frac{1}{2}\log\left(z_{s}\right)+\frac{1}{2}\log\left(z_{l}\right)=\log\left(z_{b}\right).\label{eq:F_k=00003Dinf}\end{equation}
In Ref.\cite{BKM2011} this case was analyzed and the transition was 
found to be of order $\left\lceil \frac{c-1}{c-2}\right\rceil$, which diverges as $c\rightarrow 2^+$, 
decreases as $c$ increases and yields a $2^{nd}$ order transition for $c\ge 3$. This can be seen by expanding 
Eqs.(\ref{eq:dFdzs},\ref{eq:dFdzl},\ref{eq:case_k=00003Dinf}) near
the critical temperature, where $sz_{l}=1$. Setting $t\equiv T_{c}-T$,
$\delta m_{b}\equiv m_{b}-m_{bc}$, $\delta z_{l}\equiv z_{l}-s^{-1}$
and $\delta z_{s}=z_{s}-z_{sc}$, where $m_{bc}$ and $z_{sc}$ are the
values of $m_{b}$ and $z_{s}$ at $T_{c}$, and using the identity
$\Phi_{c}(1-\delta)\approx\zeta_{c}-\delta^{c-1}$ yields below the critical temperature 
($t>0$)
\[
\delta m_{b}\sim\delta z_{s}\sim\delta z_{l}^{c-2}\sim t ~.
\]
Hence $\delta z_{l}\sim t^{\frac{1}{c-2}}$. Expanding
Eq.(\ref{eq:F_k=00003Dinf}) to appropriate order in $t$ and $\delta
z_{l}$ yields
\begin{eqnarray}
F & = & F\left(T_{c}\right)+\alpha
t+\beta\delta z_{l}^{c-1}+O\left(t^{2}\right)\nonumber \\ & \sim &
F\left(T_{c}\right)+\alpha t+\beta
t^{\frac{c-1}{c-2}}+O\left(t^{2}\right)\label{eq:k=00003Dinf_fe_expansion}
\end{eqnarray}
Above the critical temperature ($t<0$) $\delta z_l=0$ and hence 
the $\left\lceil \frac{c-1}{c-2}\right\rceil$-th derivative of $F$ diverges as temperature approaches $T_c$, constituting a phase transition 
of the same order.  The fact that the transition becomes more pronounced (of
lower order) as the loop exponent $c$ increases
can be appreciated by inspecting Fig.\ref{fig:mb_vs_T} which shows $m_{b}$ as
function of temperature for $c$ below and above $3$.


%

\subsection{$0<\kappa<\infty$ }

In this case, as for $\kappa=0$, both $m_{b}$ and $m_{s}$ are set by
minimizing the Landau free energy given in Eq.(\ref{eq:redF_def}).
Here, however, there is no simple relation between the
fugacities\begin{eqnarray} 0&=&\frac{\partial F}{\partial
  m_{b}}\nonumber \\ & = &
\log\left[\frac{z_{b}}{z_{l}}\right]-\beta\kappa\frac{\left(1-m_{b}-2m_{s}\right)\left(1+m_{b}\right)}{\left(m_{b}+m_{s}\right)^{2}}\label{eq:dFdmb}\\ 0&=&\frac{\partial
  F}{\partial m_{s}} = \log\left[\frac{z_{s}}{z_{l}}\right] +
\nonumber
\\ &&-\beta\kappa\frac{\left(1-m_{b}-2m_{s}\right)\left(1+3m_{b}+2m_{s}\right)}{\left(m_{b}+m_{s}\right)^{2}}\label{eq:dFdms}\end{eqnarray}
These equations, together with Eqs.(\ref{eq:zs_sol},\ref{eq:zl_eq})
for $T\le T_c$, and Eqs.(\ref{eq:zs_sol}) and $z_{l}=s^{-1}$
for $T\ge T_c$ , set the value of the order
parameter $m_{b}$ in the thermodynamic limit. Inserting Eqs.(\ref{eq:dFdmb},\ref{eq:dFdms}) into
Eq.(\ref{eq:F_def}) yields
\begin{equation*}
F(T) = \log\left[z_{b}\left(m_b,m_s\right)\right]-\beta\kappa\frac{\left(1-m_{b}-2m_{s}\right)^{2}}{\left(m_{b}+m_{s}\right)^{2}}.
\end{equation*}
Repeating argument used for $\kappa=\infty$ shows that here, too, the order of
the transition is $\left\lceil \frac{c-1}{c-2}\right\rceil$.

We observe in Fig.\ref{fig:mbms_plane} that the trajectories for $\kappa=0,1,\infty$ in the $(m_b,m_s)$ plane intersect at 
a single point $\left(m_{b}^{*},m_{s}^{*}\right)$. In fact, this special point is common to 
all such trajectories with arbitrary $\kappa$: Let 
$T^{*}$ be the temperature for which the minimum of the free energy
$F\left(m_{b},m_{s}\right)$ satisfies
$m_{s}^{*}=\frac{1-m_{b}^{*}}{2}$ for some $\kappa$. Then, for any
other $\kappa$, the minimum of the free energy at $T^*$ is also given by 
$\left(m_{b}^{*},m_{s}^{*}\right)$, because 
the $\kappa$-dependent part of the free energy
$\kappa\frac{\left(1-m_{b}-2m_{s}\right)^{2}}{m_{b}+m_{s}}$ vanishes (and hence is minimal)
when $m_{s}=\frac{1-m_{b}}{2}$.

\begin{figure}
\includegraphics[scale=0.5]{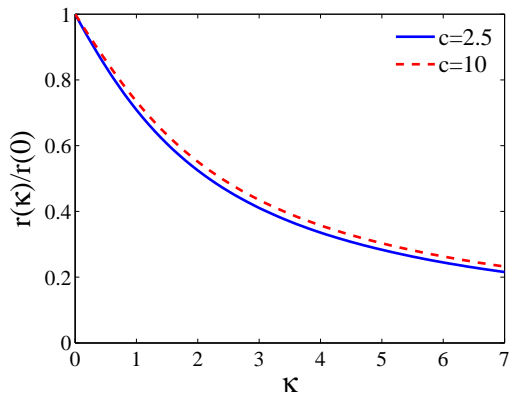}\caption{\label{fig:rkappa}$r(\kappa)/r(0)$, where $r\equiv1-m_{bc}-2m_{sc}$,
for different values of $c$. It can be seen that $r(\kappa)/r(0)$ depends only
weakly on $c$, and that, as expected, it decays monotonically from
$1$ for $\kappa=0$ to $0$ at $\kappa=\infty$.}

\end{figure}

We now consider the overtwist linking number at criticality for $0<\kappa<\infty$. This number, $r=1-m_{bc}-2m_{sc}$, cannot be 
obtained analytically. In Fig.\ref{fig:rkappa} we present the
numerically calculated $r(\kappa)/r(0)$ ratio for two values of $c$. $r(\kappa)$ depends weakly on $c$ and could be either positive or negative,
depending on the parameters of the model. However, for a given set of parameters, the 
sign of $r(\kappa)$ does not change with $\kappa$. In order to demonstrate this point, consider the 
special point in Fig.\ref{fig:mbms_plane} where all trajectories for different $\kappa$ intersect at a shared temperature $T^*$ and note that this point is the borderline between negative and positive $r(\kappa)$ on each trajectory. 
Therefore, if the parameters are such that the intersection is to the left of the singular line $\Gamma$, then for a given $\kappa$ the critical temperature satisfies $T_c<T^{*}$ and hence $r(\kappa)<0$. If, on the other hand, the intersection is to the right of the singular line then $r(\kappa)>0$ for the same reason. In addition, if the parameters are such that the intersection is to the left of the singular line, a corollary follows that $T^{*}=T_{c}^{(\kappa=0)}$, which in turn implies $T_{c}^{(\kappa>0)}<T_{c}^{(\kappa=0)}$. Recalling that $\kappa=0$ refers to the case with no LK conservation, this is in agreement with the experimental evidence that imposing circular topology reduces the melting temperature \cite{VAAP2000}. There is no equivalent statement in the other case in which the intersection is to the right of the singular line. 

\section{Conclusions}

In this paper we analyzed the thermal denaturation of a circular
DNA molecule, in which the linking number is conserved. Within the
framework of the Poland-Scheraga model, we have considered the
two possible mechanisms for conserving the LK: writhing (forming supercoils) and (over)twisting.
The denaturation transition is studied for arbitrary values of the elastic constant $\kappa$ associated with the overtwist elastic energy. We found that the model exhibits no transition for $c \le 2$ and a high-order, continuous transition 
for $c>2$, $\kappa>0$. 
The singular part of the free energy was found to scale as $t^\frac{c-1}{c-2}$
with $t=T_c-T$, yielding a transition of order $\left\lceil
\frac{c-1}{c-2}\right\rceil$. The order of the transition diverges as $c$ approaches 2 from above,
it decreases with increasing $c$ and it becomes second order for $c\ge 3$.
The model with $\kappa=0$ behaves differently, exhibiting no transition for $c \le 2$
and a first order transition for $c>2$. Similar observations were reported before
for the limiting cases restricted to supercoils only \cite{BKM2011} and
overtwist only \cite{KBM2012}.

The canonical analysis carried out here brings new insights.
For example, the first-order transition which takes place for $\kappa=0$ and $c>2$
is found to be rather special in that it does not have a metastable region (see
Fig.\ref{fig:F_vs_mb}a).  This is true also for the
original PS model. In addition, the analysis of the $(m_b,m_s)$ trajectories unveiled a $\kappa$-independent special point $\left(m_{b}^{*},m_{s}^{*}\right)$ which in return led to the prediction that $T_{c}^{(\kappa>0)}<T_{c}^{(\kappa=0)}$ (the melting temperature reduced by circular topology) for a wide range of parameters, in line with an earlier experimental observation. 

The model considered in this paper corresponds to a homogeneous circular DNA chain, while biological DNA 
molecules are heterogeneous. However, through the Harris criterion
\cite{Har1974} we find that the disorder is irrelevant for $\kappa>0$ and $c<3$, where the specific heat exponent $\alpha=2-\frac{c-1}{c-2}=\frac{c-3}{c-2}$ is negative. Therefore we do not expect 
the sequence heterogeneity to change the nature of the phase transition and the associated critical exponents. 
As the actual value of the loop exponent was estimated to be $c\approx2.12$
\cite{KMP2000}, our analysis should be valid for sufficiently long,
real DNA chains.

Previous accounts on denaturation of circular DNA have found that a
macroscopic loop is formed above $T_{c}$, reminiscent of Bose-Einstein
condensation. Although not discussed here, we expect a similar
phenomenon in the combined model of supercoils and overtwist, and it would be
be interesting to analyze the linking number exchange between the
macroscopic loop, the microscopic loops, and the supercoiled and
overtwisted segments.

{\bf Acknowledgements:} We thank O. Cohen, O. Hirschberg, S. Medalion and Y. Rabin for helpful discussions. This work was supported by the Israel Science Foundation (ISF) and the Turkish Technological and Scientific Research
Council (TUBITAK) through the grant TBAG-110T618.

\section{Appendices}

\subsection{Appendix A: Monotonicity of the RHS of Eq.(\ref{eq:zl_eq})}
We wish to show that \[
f\left(z_{l}\right)=\frac{A\Phi_{c-1}(sz_{l})}{\frac{1}{1-\nu
    z_{s}}+A\Phi_{c}(sz_{l})}\] is a monotonically increasing function
for $z_{l}\in(0,s^{-1})$.  Using Eq.(\ref{eq:zs_sol}) we can
write\begin{eqnarray*} g\left(\eta,x\right) & \equiv\frac{1}{\eta} &
\frac{1}{1-\nu z_{s}\left(z_{l},x\right)}\\ & = &
\frac{2}{\sqrt{\left(1+x\right)^{2}+4x\eta}-\left(1+x\right)}\end{eqnarray*}
where $\eta=A\Phi_{c}\left(sz_{l}\right)$ increases and
$g\left(\eta,x\right)$ decreases with $z_{l}$. Hence we can write \[
f\left(z_{l}\right)=\frac{1}{1+g\left(\eta,x\right)}\times\frac{\Phi_{c-1}(sz_{l})}{\Phi_{c}(sz_{l})}\]
where the first factor is an increasing function of $z_{l}$. It is thus
sufficient to show that the second factor also increases with
$z_{l}$.  To this end, we differentiate this term: \begin{eqnarray*}
  \frac{d}{dz_{l}}\left[\frac{\Phi_{c-1}(sz_{l})}{\Phi_{c}(sz_{l})}\right]
  & = &
  \left[\frac{\Phi_{c-2}(sz_{l})\Phi_{c}\left(sz_{l}\right)}{\Phi_{c}(sz_{l})^{2}}\right.\\ &
    &
    \left.-\frac{\Phi_{c-1}(sz_{l})^{2}}{\Phi_{c}(sz_{l})^{2}}\right].\end{eqnarray*}
Now we show that the numerator of the derivative, denoted by
$\Sigma\left(z_{l}\right)$ is positive, by expressing the
polylogarithm function explicitly as a power series of the variable
$y=sz_{l}$:\begin{eqnarray*} \Sigma\left(z_{l}\right) & = &
  \sum_{k,l=1}^{\infty}\left[\frac{y^{l}}{l^{c-2}}\frac{y^{k}}{k^{c}}-\frac{y^{l}}{l^{c-1}}\frac{y^{k}}{k^{c-1}}\right]\\ &
  = &
  \sum_{k,l=1}^{\infty}\frac{y^{l}}{l^{c-1}}\frac{y^{k}}{k^{c}}\left[l-k\right]\\ &
  = &
  \sum_{k<l}^{\infty}y^{l+k}\left[l-k\right]\left[\frac{1}{l^{c-1}}\frac{1}{k^{c}}-\frac{1}{k^{c-1}}\frac{1}{l^{c}}\right]\\ &
  = &
  \sum_{k<l}^{\infty}\frac{y^{l+k}}{l^{c}k^{c}}\left[l-k\right]^{2}>0\ .\end{eqnarray*}
This demonstrates that $f(z_l)$ is a monotonically increasing function
of $z_l$.

\subsection{Appendix B: Branch-cut integration}

\begin{figure}
\includegraphics[width=0.4\textwidth]{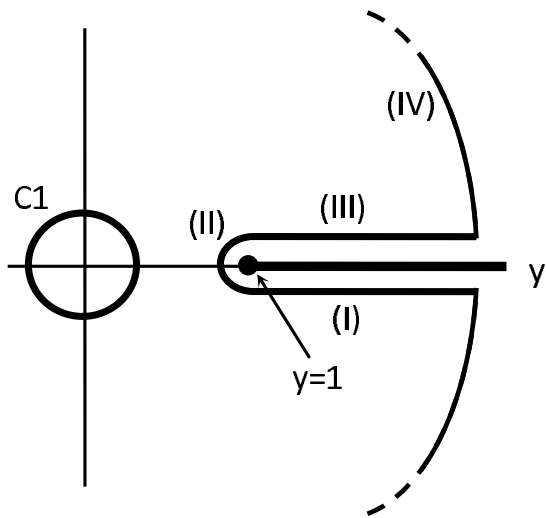}

\caption{\label{fig:contour}The original contour of integration C1 can be
deformed to the contour which is given by Eq.(\ref{eq:contour_def}) and is described in the text of Appendix B. The branchpoint $y=1$ is marked here
by a filled circle.}

\end{figure}

We wish to evaluate the integral for the partition function with
$\kappa=0$ given in Eq.(\ref{eq:Z_fe}) \[
\psi\left(L_{b},L_{s},L_{l}\right)=\frac{1}{2\pi i}\oint
e^{-L\tilde{F}_{\kappa=0}\left(z_{s},z_{l},m_{b},m_{s}\right)}dz_{l},\]
with $z_{s}\left(x,z_{l}\right)$ given by Eq.(\ref{eq:zs_sol}).
$L_{i}$ satisfy $L_{b}+L_{l}+L_{s}=L$, $L_{b}/L_{s}=x$ and
$m_{i}=L_{i}/L$ ($i=b,s$). Defining $y=sz_{l}$ yields
\begin{eqnarray*}
\psi\left(L_{b},x\right) & = & \frac{s^{L_{l}}}{2\pi i}\oint\frac{\left[\frac{1}{1-\nu z_{s}}+A\Phi_{c}\left(y\right)\right]^{L_{b}}}{z_{s}^{L_{b}/x}y^{L-L_{b}\frac{1+x}{x}}}dy\\
 & = & \frac{s^{L_{l}}}{2\pi i}\oint I\left(y\right)^{L}dy.\end{eqnarray*}
Eq.(\ref{eq:mb_c_vs_x}) defines the value of $m_{b}=L_{b}/L$ below
which the integrand has no saddle point, therefore the integral should be
evaluated by another method. The integration contour can be deformed
to the contour depicted in Fig.\ref{fig:contour}, composed of the
following segments: \begin{eqnarray}
(I) & : & \left[R-i\epsilon,1-i\epsilon\right]\nonumber \\
(II) & : & \left\{ 1-\epsilon e^{i\theta}:\frac{\pi}{2}<\theta<\frac{3\pi}{2}\right\} \nonumber \\
(III) & : & \left[1+i\epsilon,R+i\epsilon\right]\nonumber \\
(IV) & : & \left\{ Re^{i\theta}:\delta<\theta<2\pi-\delta\right\} \label{eq:contour_def}\end{eqnarray}
where $R\rightarrow\infty$ and $tg\left(\delta\right)=\frac{\epsilon}{R}$.
We wish to show now that the only contribution comes from the vicinity
of the branch point: To see that the contribution of $(IV)$ is negligible
we note that for $\left|y\right|\rightarrow\infty$,$\left|\Phi_{c}(y)\right|\rightarrow\frac{\log(y)^{c}}{\Gamma(c+1)}$
\cite{Lewin1981}. From Eq.(\ref{eq:zs_sol}) we see that for $\left|y\right|\rightarrow\infty$
$z_{s}\rightarrow1$ and $\frac{1}{1-\nu z_{s}}\rightarrow\sqrt{\Phi_{c}\left(y\right)}$
so that for large enough $L$ and $x<\infty$,
\[
\lim_{R\rightarrow\infty}\left[I\left(Re^{i\theta}\right)\right]^{L}\sim \lim_{R\rightarrow\infty}\frac{\log(R)^{cL_{b}}}{R^{L-L_{b}\frac{1+x}{x}}}<\lim_{R\rightarrow\infty}\frac{1}{R^{2}}\ .
\]
Hence segment $\left(IV\right)$ of the contour has no contribution. Along segment $(II)$
the function $I(y)$ is analytic and hence the integral is of order
$\epsilon$ and can be taken to be arbitrarily small.

$I(y)$ can be written as a power series with only real coefficients,
and hence $I\left(y^{*}\right)^{L}=I^{*}\left(y\right)^{L}$, where
$y^{*}$ is the complex conjugate of $z$. Integrating along segments
$(I)+(III)$ yields\begin{eqnarray*} \psi & = & \frac{s^{L_{l}}}{2\pi
  i}\int_{(I)+(III)}I\left(y\right)^{L}dy\\ & = &
\frac{s^{L_{l}}}{\pi}\int_{1}^{\infty}Im\left[I\left(y\right)^{L}\right]dy.\end{eqnarray*}
Defining $I(y)=\Lambda(y)e^{i\Psi(y)}$, where $\Lambda(y)$ and
$\Psi(y)$ are real functions, we can write
$Im\left[I(y)^{L}\right]=$$\Lambda(y)^{L}sin\left[L\Psi(y)\right]$.
As $\Psi(y)$ is a smooth function for $y>1$ in the thermodynamic limit
the oscillations in the $\sin()$ function average out to zero.
Therefore, the only contribution to the integral comes from the
vicinity of $y=1$, where $\Psi(y)$ has a discontinuity in some
derivative.

The function $I(y)$ has a pole at $y=0$ where
$I(y\rightarrow0_{+})\rightarrow\infty$, so that $I'(y)<0$ near the
origin. When $c>2$, the fact that there is no saddle point for $0<y<1$
implies that $I'(1)<0$ as well, and hence $\Lambda'(1)<0$. For $\delta
y\equiv y-1\ll1$ the imaginary part of the polylogarithm function is
approximately $Im\left[\Phi_{c}(1+\delta y)\right]\approx a\delta
y^{c-1}$ with $a=\pi/\Gamma(c)$, and therefore $\Psi(1+\delta
y)\approx\tilde{a}\delta y^{c-1}$ with
$\tilde{a}=\left(a/\Lambda(1)\right)\times\partial
I/\partial\Phi_{c}(y)$.  Combining these observations yields to
leading order $I(1+\delta y)\approx\Lambda(1)\exp\left[-b\delta
  y+i\tilde{a}\delta y^{c-1}\right]$ with
$b=-\Lambda'(1)/\Lambda(1)>0$. \begin{eqnarray*} \psi & \approx &
  \frac{s^{L_{l}}}{\pi}\Lambda(1)^{L}\int_{0}e^{-bL\delta
    y}\sin\left(L\tilde{a}\delta y^{c-1}\right)d\delta
  y.\end{eqnarray*} The contribution to the integral comes from a
region of size $\delta y\sim\frac{1}{L}$, so that the upper limit can
be stretched to $\infty$ without affecting the result. Rescaling by
$\tilde{y}=bL\delta y$ we obtain
\begin{eqnarray*}
\psi & \approx &
  \frac{s^{L_{l}}}{\pi}\Lambda(1)^{L}\left(bL\right)^{-1}\int_{0}^{\infty}e^{-\tilde{y}}\sin\left(\frac{\tilde{a}\tilde{y}^{c-1}}{b^{c-1}L^{c-2}}\right)d\tilde{y}\ .
\end{eqnarray*}
As $c>2$, in the thermodynamic limit $L\rightarrow\infty$ the argument
of the $\sin()$ function is small and it can be
expanded.\begin{eqnarray*} \psi & \approx &
  \frac{\tilde{a}\left(bL\right)^{c-2}}{\pi
    L}s^{L_{l}}\Lambda(1)^{L}\int_{0}^{\infty}e^{-\tilde{y}}\tilde{y}^{c-1}d\tilde{y}\\ &
  = & \frac{\tilde{a}\left(bL\right)^{c-2}\Gamma\left(c\right)}{\pi
    L}s^{L_{l}}\Lambda(1)^{L}\ .
\end{eqnarray*}
Substituting back $z_{l}=y/s$ yields \[
\psi\left(L_{b},L_{s},L_{l}\right)\approx\frac{\tilde{a}\left(bL\right)^{c-2}\Gamma\left(c\right)}{\pi
  L}\,e^{-L\tilde{F}_{\kappa=0}\left(z_{l}=s^{-1},z_{s},m_{b},m_{s}\right)},\]
where $z_{s}=z_{s}\left(z_{l},x\right)$ is given in
Eq.(\ref{eq:zs_sol}). Hence up to logarithmic corrections the free
energy is given by its value at the branch-point.

\bibliography{supercoil2}
\bibliographystyle{h-physrev}

\end{document}